\documentclass[]{interact}

\usepackage{epstopdf}% To incorporate .eps illustrations using PDFLaTeX, etc.
\usepackage[caption=false]{subfig}% Support for small, `sub' figures and tables

\theoremstyle{plain}% Theorem-like structures provided by amsthm.sty

\theoremstyle{definition}

\theoremstyle{remark}

\usepackage[numbers,sort&compress]{natbib}
\bibpunct[, ]{[}{]}{,}{n}{,}{,}
\renewcommand\bibfont{\fontsize{10}{12}\selectfont}

\usepackage{lineno,hyperref}
\usepackage{color}
\usepackage{graphicx}

\providecommand{\bo}{\mathbf}
\providecommand{\bs}{\boldsymbol}

\providecommand{\tr}{\mbox{tr}}

\begin{document}

\title{Robust multi-outcome regression with correlated covariate blocks using fused LAD-lasso}
% Robust multi-outcome fused LAD-lasso regression with correlated covariance blocks: application to retirement behavior study
% Robust multi-outcome regression with correlated covariance blocks using fused LAD-lasso: application to retirement data

\author{
\name{Jyrki M\"ott\"onen\textsuperscript{a}
\thanks{
CONTACT Jyrki M\"ott\"onen. Email: jyrki.mottonen@helsinki.fi\\ }
%\hspace*{1.8em}Preprint of this manuscript is available in arXiv [x]},
Tero L\"ahderanta\textsuperscript{b},
Janne Salonen\textsuperscript{c}
and
Mikko J. Sillanp\"a\"a\textsuperscript{b}
}
\affil{\textsuperscript{a}Department of Mathematics and Statistics, University of Helsinki, Helsinki, Finland; \textsuperscript{b}Research Unit of Mathematical Sciences, University of Oulu, Oulu, Finland;\\
\textsuperscript{c}Finnish Public Sector Pension Provider Keva, Helsinki, Finland}
}

\maketitle

\begin{abstract}
Lasso is a popular and efficient approach to simultaneous estimation and variable selection in high-dimensional regression models. In this paper, a robust LAD-lasso method for multiple outcomes is presented that addresses the challenges of non-normal outcome distributions and outlying observations. Measured covariate data from space or time, or spectral bands or genomic positions often have natural correlation structure arising from measuring distance between the covariates. The proposed multi-outcome approach includes handling of such covariate blocks by a group fusion penalty, which encourages similarity between neighboring regression coefficient vectors by penalizing their differences for example in sequential data situation. Properties of the proposed approach are first illustrated by extensive simulations, and secondly the method is applied to a real-life skewed data example on retirement behavior with  heteroscedastic explanatory variables. %are highly correlated.

%Lasso is a popular and efficient approach for simultaneous estimation and variable selection in high-dimensional regression models. A robust LAD-lasso method is presented here for multiple outcomes, which may have non-normal distributions or contain outlying observations. Measured covariate data made in space or time, or in spectral bands or genomic positions often have natural correlation structure arising from measuring distance between covariates. The proposed multi-outcome approach includes handling of such covariate blocks by a group fusion penalty, which encourages similarity between neighboring regression coefficient vectors by penalizing their differences. Properties of the proposed approach are illustrated by simulations and a real-world data example on retirement behavior.
\end{abstract}


%%%%%%%%%%%%%%%%%%%%%%%%%%%%%%%%%%%%%%%%%%%%%%%%%%%%%%%%%%%%%%
\section{Introduction}
\label{sec:intro}

In high-dimensional regression problems, when the number of predictors is much higher than data points ($p \gg n$), it is common to use sparsity-inducing variable selection for choosing important predictors into the model. Such approaches ideally operate by inducing a heavy shrinkage on coefficients of spurious (unimportant) predictors towards zero while true coefficients are experiencing practically no shrinkage at all. By following such principle, \textit{least absolute shrinkage and selection operator} (\textit{lasso}) is a popular tool for simultaneous estimation and variable selection in single-outcome \cite{tibshirani1996,li2012} and in multi-outcome \cite{turlach2005, yuan2006, yuan2007} settings. 

Robust methods are needed in cases when the data sample follows skewed distribution or contain some amount of outlying 
observations, as well as, have leverage points among covariates, which can all clearly violate the typical distributional assumptions of the model.
A general way to make model robust against outliers of outcome variable is to assume some long-tailed distribution, such as, t-distribution or Laplace distribution for the residuals \cite{lange1989,yang2009,chen2014}.  
\textit{Least absolute deviation} (\textit{LAD}) regression and lasso regression have been combined to \textit{single-outcome LAD-lasso} \cite{wang2008} and \textit{multi-outcome LAD-lasso} \cite{mottonen2015,li2015}, which are robust extensions to the ordinary lasso models.
%\textit{Single-outcome LAD-lasso} \cite{wang2008} and \textit{multi-outcome LAD-lasso} \cite{mottonen2015,li2015} have been proposed as robust extensions for the ordinary lasso models. 
If considered under the likelihood framework, these approaches correspond to assuming Laplace distribution for the residuals. Robustness against outliers of explanatory variables (i.e., leverage points) are rarely considered together with lasso. Exception being an \textit{adaptive LAD-lasso} approach, which was recently proposed to handle excess number of zeros in covariate data, see \citep{mottonen2021}, and a real data example in \citep{Lahderanta2022}.

Sometimes in high-dimensional regression problems, there is natural block-structure (or ordering) among measured covariates in such a way that nearby covariates are more correlated than farther ones. This occurs naturally for example in spatial statistics, time-series, spectroscopy, genomic data sets, and so on. 
%%\textcolor{blue}{It is currently popular to analyse longitudinal data using varying-coefficient models ().}
%%In genetics, varying-coefficient models can be used to map genes 
%%influencing to the time-course 
%%phenotypic data \citep{li2015}. 
This case is generally considered a drawback in ordinary lasso regression, because it has a tendency to select only one or very few representatives out from such highly correlated blocks of covariates. To alleviate this, \textit{ridge regression} or \textit{elastic net} has been suggested to be used in such a case \citep{Zou2005}.

Alternatively, one could argue that analysis methods of such covariate block data sets could ideally apply smoothing to share information between adjacent regression coefficients. One such approach is \textit{fused lasso}, which encourages similarity between the neighboring regression coefficients by penalizing their differences \cite{tibshirani2005,tibshirani2011}. Note that one could also justify and interpret the use of fusion penalty from dimension reduction perspective \citep{zheng2015,She2020}. In fact fused lasso technique is used in various analytical contexts, such as, clustering data \cite{heinzl2014,hacking2011}, image processing \citep{tibshi2008, tibshirani2011}, network estimation \citep{danaher2014}, and in econometric applications \citep{corsaro2021}.

Closely related to our approach, there is already some previous work on the \textit{single-outcome fused LAD-lasso} approach \citep{liu2018}, and further in adaptive context \citep{wang2016}.
Multi-outcome lasso model, having own regression coefficient for each outcome, needs to be adjusted in order to consider fused lasso in multi-outcome data context.
%%in varying-coefficient model/ modelling? context.
Again, multi-outcome LAD-lasso yields a more robust alternative in the presence of heavy-tailed distribution or outlying observations 
\cite{mottonen2015,li2015}.
However, unlike in conventional multi-outcome models (e.g., in \cite{mottonen2015} and \cite{li2015}), 
the regression coefficients of the neighboring covariates (e.g., time-points) should be further smoothed (that is, to be more alike) 
to fully utilize sequential nature of the covariate data. By smoothing, the neighboring coefficients can borrow "strength" or information from each others during estimation. The smoothing property can be obtained by 
including an additional fusion penalty to the penalized regression problem. 
%\textcolor{blue}{Thus, the formulated model, which we present here is called \textit{multi-outcome fused LAD-lasso}}.

In short, the challenge we face in this paper is variable selection with correlated explanatory variables in the presence of heavy-tailed outcome distributions.  
The novelty of this paper is introducing the \textit{fused group LAD-lasso} in \textit{multi-outcome} context and demonstrating its desirable performance with simulations and real-life data example. We also provide a preliminary R-package with which the researchers can try out the proposed method with simulated data set or in their own applications.

%%Sometimes in lasso models it is important to take into account the dependency among individual samples / observations.
%%This means that there is some  residual structure between observations.
%%This can be accounted for by having an extra random effect in the model called \textit{mixed lasso} \cite{wang2010}
%%or having residual covariance structure in the lasso model using \textit{generalized least squares}  \cite{huang2010}.
%%On the other hand, \textit{LAD estimation} has been introduced for clustered data (e.g., \cite{nevalainen2015}), but these approaches do not consider lasso-type penalized estimation simultaneously. Our aim is to combine these two lines of research, taking into account clustering of observations in the LAD-lasso framework. 

Next, in Section \ref{sec:LAD-lasso}, we outline the already established multi-outcome group LAD-lasso method, and in Section \ref{sec:fused group LAD-lasso} we introduce the fused method in multi-outcome setting. In Section \ref{sec:simulation} are the results of a simulation experiment. In Section \ref{sec:realdata} we show how the proposed technique can handle real-world data related to retirement behaviour in $n \gg p$ situation. Finally, in Section \ref{sec:concluding} we give some closing remarks.

%%%%%%%%%%%%%%%%%%%%%%%%%%%%%%%%%%%%%%%%%%%%%%%%%%%%%%%%%%%%%%
%%%%%%%%%%%%%%%%%%%%%%%%%%%%%%%%%%%%%%%%%%%%%%%%%%%%%%%%%%%%%%
\section{Multi-outcome LAD-lasso}
\label{sec:LAD-lasso}

Consider a multi-outcome multiple regression model
\begin{eqnarray}
\label{model}
\bo Y = \bo X\bo B + \bo E, 
\end{eqnarray}
where $\bo Y = (\bo y_1,\ldots,\bo y_n)'$ is an $n\times q$ matrix of $n$ observed
values of $q$ outcome variables,
$$
\bo X = 
\begin{pmatrix}
\bo x_1' \\ 
\bo x_2' \\
\vdots\\
\bo x_n'
\end{pmatrix}
=
\begin{pmatrix}
1 & x_{11} & \cdots & x_{1p} \\ 
1 & x_{21} & \cdots & x_{2p} \\
\vdots & \vdots & \ddots & \vdots \\
1 & x_{n1} & \cdots & x_{np}
\end{pmatrix}
$$
is an $n\times(p+1)$ matrix of $n$ observed values of $p$ explaining variables,
$
\bo B = (\bs\beta_0, \bs\beta_1, \ldots, \bs\beta_p)'
$ 
is a $(p+1)\times q$ matrix of regression
coefficients ($\bs\beta_0$ is the intercept vector), and $\bo E = (\bs\varepsilon_1,\ldots,\bs\varepsilon_n)'$ is an $n\times q$ 
matrix of $q$-dimensional errors. We further assume that $\bs\varepsilon_1,\ldots,\bs\varepsilon_n$ is a random sample of 
size $n$ from a $q$-variate distribution centred at the origin.

The minimizer of the objective function
\begin{eqnarray}
\label{ols2}
&&\frac1n\sum_{i=1}^n\|\bo y_i - \bo B'\bo x_i\|^2
=
\frac1n\sum_{i=1}^n(\bo y_i  - \bo B'\bo x_i)'(\bo y_i  - \bo B'\bo x_i) \nonumber
\\
&& = 
\frac1n\tr\left[(\bo Y-\bo X\bo B)'(\bo Y-\bo X\bo B)\right]
\end{eqnarray}
gives the ordinary least squares estimate and 
when the rank of 
$\bo X$
 is $p+1$, the objective function is minimized when 
\begin{eqnarray*}
\label{ols12}
\hat{\bo B}
& = &
\left(\bo X'\bo X\right)^{-1}
\bo X'
\bo Y.
\end{eqnarray*}

The single-outcome lasso estimation method can be generalized straightforwardly to the multi-outcome version by using 
the penalized objective function
\begin{eqnarray}
\label{lasso2}
\frac1n\sum_{i=1}^n\|\bo y_i-\bo B'\bo x_i\|^2 + \lambda\sum_{j=1}^p\gamma_j\|\bs\beta_j\|.
\end{eqnarray}
where
$$
\gamma_j =
\begin{cases}
1, & \text{if the $j$th covariate is penalized,}\\
0, & \text{otherwise.}
\end{cases}
$$
and $\lambda\geq 0$ is a shrinkage parameter. When $\lambda=0$ we get the ordinary least squares estimate (if the rank of $\bo X$ is $p+1$) and 
as $\lambda$ is increased, the estimates of the coefficients shrink progressively towards zero and some may become exactly zero, see e.g. \cite{turlach2005}, \cite{yuan2006} and \cite{yuan2007}.
The minimizer of the objective function (\ref{lasso2}) now gives the {\it multi-outcome lasso estimate} for 
the regression coefficient matrix $\bo B$. 

The multi-outcome lasso method gives sparse solutions, but it is obviously not very robust. A more robust version of the model is achieved by replacing the squared norm with a L1 norm in the objective function \ref{lasso2}.
 
In other words, we minimize the LAD-lasso criterion 
\begin{equation}
\label{lad-lasso3}
\frac1n\sum_{i=1}^n\|\bo y_i-\bo B'\bo x_i\|
+ \lambda\sum_{j=1}^p\gamma_j\|\bs\beta_j\|
\end{equation}
with respect to 
$\bo B$
\cite{mottonen2015,li2015}.
We can write the objective function (\ref{lad-lasso3}) in a more compact way by realizing that 
the penalty term can be written in the same form
as the LAD term:
$$
\lambda\sum_{j=1}^p\gamma_j\|\bs\beta_j\|=
\frac{1}{n}\sum_{j=1}^p\left\|\bo 0_q - \bo B'
\begin{pmatrix}
0\\
n\lambda\gamma_j\bo e_j
\end{pmatrix}
\right\|,$$
where
$\bo 0_q$ is a $q\times1$ vector of zeros and 
$\bo e_j$ is a $p\times1$ unit norm vector with the $j$th element equal to 1 and zeros elsewhere. 
Let $j_1, j_2, \ldots, j_d$ denote the indices of the penalized explaining variables, i.e. $\gamma_{j}=1 \Leftrightarrow j\in\{j_1, j_2, \ldots, j_d\}$. 
We can now define 
\begin{eqnarray*}
\bo Y_* = 
\begin{pmatrix}
\bo y_{*,1}'\\
\vdots\\
\bo y_{*,n+d}'
\end{pmatrix}
=
\begin{pmatrix}
\bo Y\\
\bo O_{d\times q}
\end{pmatrix}
\ \ \text{and}\ \ 
\bo X_* =
\begin{pmatrix}
\bo x_{*,1}'\\
\vdots\\
\bo x_{*,n+d}'
\end{pmatrix}
=
\begin{pmatrix}
\bo X\\
n\lambda\bo A
\end{pmatrix},
\end{eqnarray*}
where $\bo O_{p\times q}$ is a $p\times q$ matrix of zeros and 
$$
\bo A = 
\begin{pmatrix}
0 & \bo e_{j_1}' \\ 
\vdots & \vdots \\
0 & \bo e_{j_d}'
\end{pmatrix}
$$
is a $d\times (p+1)$ matrix.
Then the objective function (\ref{lad-lasso3}) reduces to LAD estimation objective function \cite{oja2010}
\begin{equation}
\label{lad-lasso4}
\frac1{n+d}\sum_{i=1}^{n+d}\|\bo y_{*,i}-\bo B'\bo x_{*,i}\|,
\end{equation}
which shows that we can use any multivariate LAD regression estimation routine to find the multi-outcome LAD-lasso estimate. It is possible, for example, to use the 
function {\it mv.l1lm} of R-package {\it MNM} 
\cite{nordhausen2011,nordhausen2016}. Note, that when fitting the model using the objective function (\ref{lad-lasso4}) there is no intercept term because
the first column of $\bo X_*$ is 
$\begin{pmatrix}
\bo 1_n' & \bo 0_d'
\end{pmatrix}'$.

%\vfill\eject 
%%\section{Extension of group fusion penalty to multi-outcome situation }
%%A fusion penalty used in standard fused lasso is calculated as $\|\beta_j-\beta_{j-1}\|=\sum_{j=1}^{p-1} |\beta_{j+1}-\beta_{j}|$ while its group penalty version is here specified as $\|\bs\beta_j-\bs\beta_{j-1}\|= \sqrt{\sum_{k=1}^q(\beta_{j,k}-\beta_{j-1,k})^2}$. 

\section{Multi-outcome fused group LAD-lasso}
\label{sec:fused group LAD-lasso}

The multi-outcome LAD-lasso penalty encourages sparsity of the coefficient vectors $\bs\beta_j$. In certain empirical applications, it may be beneficial to also penalize the differences of the adjacent coefficient vectors $\bs\beta_j-\bs\beta_{j-1}$ using the so called \textit{group fusion penalty}. The term “fusion penalty” was first used by \cite{land1996}, and the single-outcome fused lasso method where the ordinary lasso is generalized by adding a fusion penalty term into the objective function was introduced by \cite{tibshirani2005}. 

Next, it is assumed that the continuous explaining variables $\bo x_1,\ldots,\bo x_p$ are ordered in a meaningful way, so that, the strongest correlations take place between neighboring variables. The multi-outcome fused group LAD-lasso estimator minimizes the following objective function with respect to 
$\bo B$:
\begin{equation}
\label{fused-lad-lasso}
\frac{1}{n}\sum_{i=1}^n\|\bo y_i - {\bo B}'\bo x_i\|
+ \lambda_1\sum_{j=1}^p\gamma_j\|\bs\beta_j\|
+ \lambda_2\sum_{k=1}^{p-1}\delta_k\|\bs\beta_{k+1}-\bs\beta_{k}\|,
\end{equation}
where 
$\lambda_2\sum_{k=1}^{p-1}\delta_k\|\bs\beta_{k+1}-\bs\beta_{k}\|=\lambda_2\sum_{k=1}^{p-1}\delta_k \sqrt{\sum_{l=1}^q(\beta_{k+1,l}-\beta_{k,l})^2}$ is the group fusion penalty, $\bs\beta_k=(\beta_{k,1},..,\beta_{k,q})'$ and
$$
\delta_k =
\begin{cases}
1, & \text{if the difference $\bs\Delta_k=\bs\beta_{k+1}-\bs\beta_{k}$ is penalized,}\\
0, & \text{otherwise.}
\end{cases}
$$
The objective function (\ref{fused-lad-lasso}) has two tuning parameters, $\lambda_1 \geq 0$ and $\lambda_2 \geq 0$. The tuning parameter
$\lambda_1$ controls the level of sparsity in the coefficient vectors $\bs\beta_{k}$ and $\lambda_2$ controls the level of sparsity in the differences of coefficient vectors
$\bs\beta_{k+1}-\bs\beta_{k}$.

In the same way as in the multi-outcome group LAD-lasso model, we can define
\begin{eqnarray*}
\bo Y_* = 
\begin{pmatrix}
\bo y_{*,1}'\\
\vdots\\
\bo y_{*,n+d_1+d_2}'
\end{pmatrix}
=
\begin{pmatrix}
\bo Y\\
\bo O_{(d_1+d_2)\times q}
\end{pmatrix}
\end{eqnarray*}
and
\begin{eqnarray*}
\bo X_* =
\begin{pmatrix}
\bo x_{*,1}'\\
\vdots\\
\bo x_{*,n+d_1+d_2}'
\end{pmatrix}
=
\begin{pmatrix}
\bo X\\
n\lambda_1\bo A_1\\
n\lambda_2\bo A_2\bo W
\end{pmatrix},
\end{eqnarray*}
where
\begin{eqnarray*}
\bo W  & = & 
\begin{pmatrix}
\bo O_{(p-1)\times 2} & \bo I_{p-1}
\end{pmatrix}
-
\begin{pmatrix}
\bo 0_{p-1} & 
\bo I_{p-1} &  \bo 0_{p-1}
\end{pmatrix} \\
& = &
\begin{pmatrix}
0 & -1       & ~~1      & 0      & \cdots & ~~0 & 0\\
0 & ~~0      &  -1      & 1      & \cdots & ~~0 & 0\\
\vdots & ~~\vdots & ~~\vdots & \vdots & \ddots & ~~0 & 0\\
0 & ~~0      & ~~0      & 0      & \cdots &  -1 & 1 
\end{pmatrix},
\end{eqnarray*}
$$
\bo A_1 = 
\begin{pmatrix}
0 & \bo e_{j_1}'\\
\vdots & \vdots\\
0 & \bo e_{j_{d_1}}'
\end{pmatrix},\ \ 
\bo A_2 = 
\begin{pmatrix}
0 & \bo e_{k_1}'\\
\vdots & \vdots\\
0 & \bo e_{k_{d_2}}'
\end{pmatrix}
$$
and
$k_1, k_2, \ldots, k_{d_2}$ denote the indices of the penalized differences $\bs\Delta_k = \bs\beta_{k+1}-\bs\beta_{k}$. 
The objective function (\ref{fused-lad-lasso}) can now be reduced to a LAD estimation objective function
\begin{equation}
\label{fused-lad-lasso2}
\frac1{n+d_1+d_2}\sum_{i=1}^{n+d_1+d_2}\|\bo y_{*,i}-\bo B'\bo x_{*,i}\|
\end{equation}
and so we can use any multivariate LAD regression estimation routine to find the multi-outcome fused LAD-lasso estimate.

\section{Simulation study}
\label{sec:simulation}

In this section, we present some experiments with simulated data sets to provide insights to the multi-outcome fused LAD-lasso in certain selected scenarios. There is also an accompanying preliminary R-package available on GitHub to test the proposed technique \footnote{https://github.com/jymottonen/mladlasso}.

For this study, we generated a data set by using a multivariate multiple regression model
$$
\bo Y = \bo X\bo B + \bo E,
$$
where $\bo Y$ is a $200\times 2$ matrix of bivariate responses, $\bo X$ is a $200\times 51$ model matrix, $\bo B = (\bs\beta_0,\bs\beta_1,\ldots,\bs\beta_{50})'$ is a $51\times 2$ matrix of regression coefficients and 
$\bo E = (\bs\varepsilon_1,\cdots,\bs\varepsilon_{200})'$ 
is a
$200\times 2$ matrix of independent bivariate errors $\bs\varepsilon_i$
distributed as
$$
\bs\varepsilon_i\sim N\left(
\begin{pmatrix}
0\\
0
\end{pmatrix},\ 
\begin{pmatrix}
1.0 & 0.7\\
0.7 & 1.0
\end{pmatrix}
\right).
$$

The values of the selected explaining variables
$\{x_{11},\ldots,x_{15}\}$ and 
$\{x_{21},\ldots,x_{25}\}$ were generated
from Gaussian AR(1) model with autoregressive 
parameters $\phi=0.9$ and $\phi=0.5$, respectively.
This indicates that the neighboring explaining variables in those two groups are then  correlated and therefore
$\text{Cor}(x_{11},x_{12})=\cdots=
\text{Cor}(x_{14},x_{15})=0.9$
and
$\text{Cor}(x_{21},x_{22})=\cdots=
\text{Cor}(x_{24},x_{25})=0.5$.
The correlation decreases towards 0 as the distance (lag) between variables increases.
All the other values of the explaining variables were 
generated from standard normal distribution (Gaussian AR(0) model).

The regression coefficient vectors were set as following:
$$ 
\bs\beta_{5} = \bs\beta_{40} = 
\begin{pmatrix}
7\\
8
\end{pmatrix},\ 
\bs\beta_{11} = \cdots = \bs\beta_{15} = 
\begin{pmatrix}
10\\
12
\end{pmatrix},\
\bs\beta_{21} = \cdots = \bs\beta_{25} = 
\begin{pmatrix}
8\\
6
\end{pmatrix}
$$
and all the other regression coefficients were
generated from standard normal distribution.

\begin{figure}[ht!]
\centering
\subfloat[LAD estimates ($\lambda_1=\lambda_2=0$).]{%
\resizebox*{14cm}{!}{\includegraphics[width=1\textwidth]{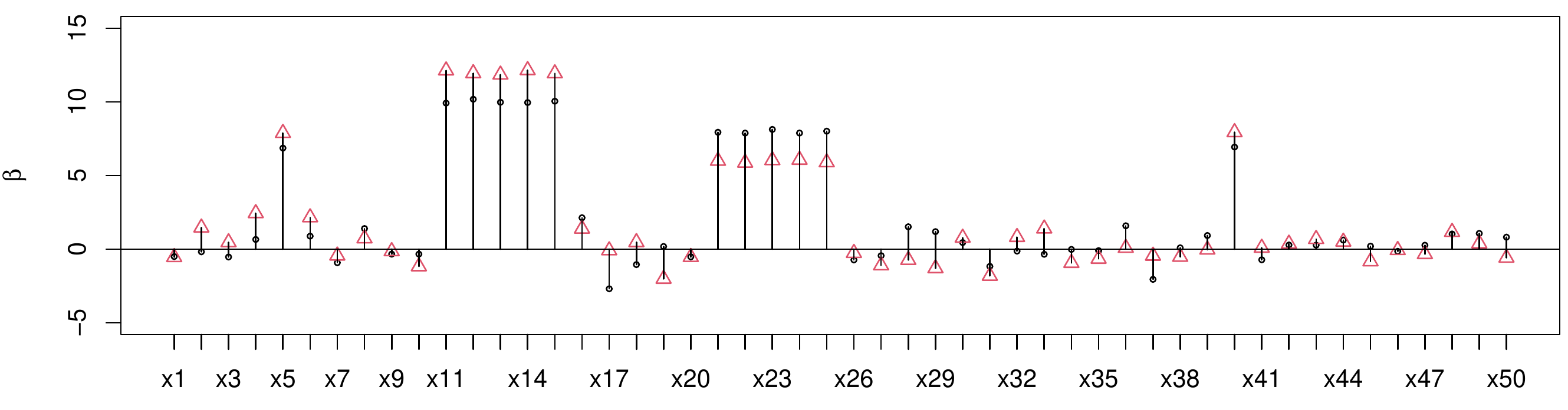}}
\label{fig:0.0}
}\\
\subfloat[LAD-lasso estimates ($\lambda_1=0.2, \lambda_2=0$).]{%
\resizebox*{14cm}{!}{\includegraphics[width=1\textwidth]{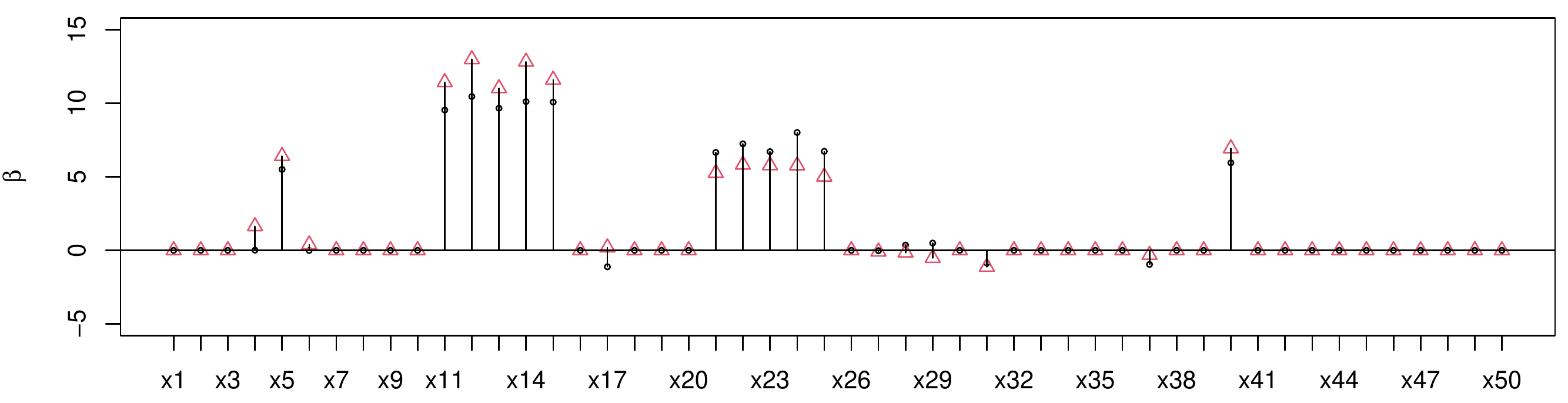}}
\label{fig:2.0}
}\\
\subfloat[Fused LAD-lasso estimates ($\lambda_1=0.2, \lambda_2=0.1$).]{%
\resizebox*{14cm}{!}{\includegraphics[width=1\textwidth]{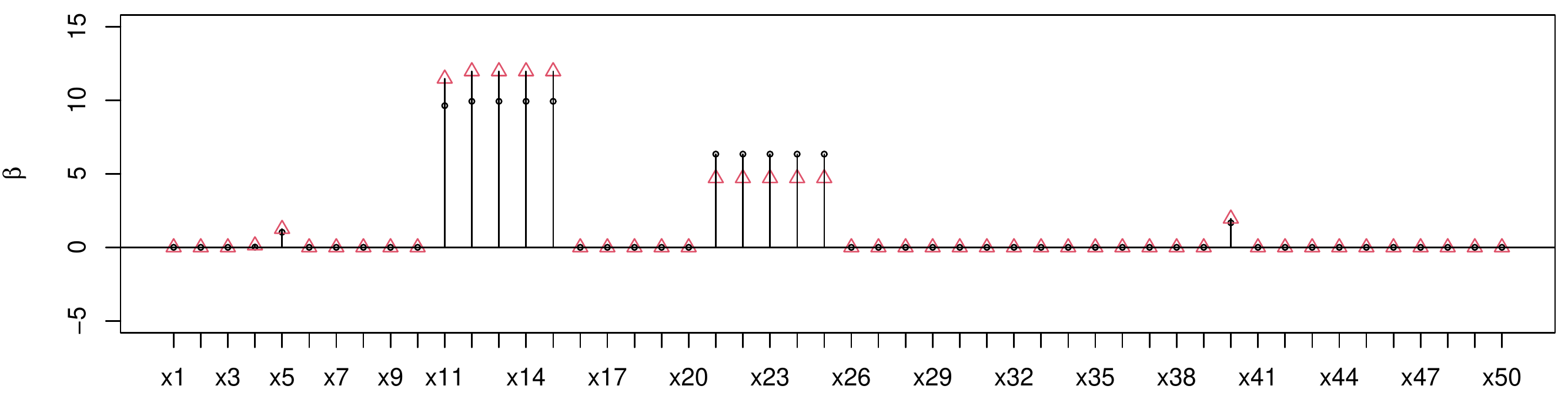}}
\label{fig:2.1}
}\\
\subfloat[Fused LAD-lasso estimates ($\lambda_1=0.2, \lambda_2=0.2$).]{%
\resizebox*{14cm}{!}{\includegraphics[width=1\textwidth]{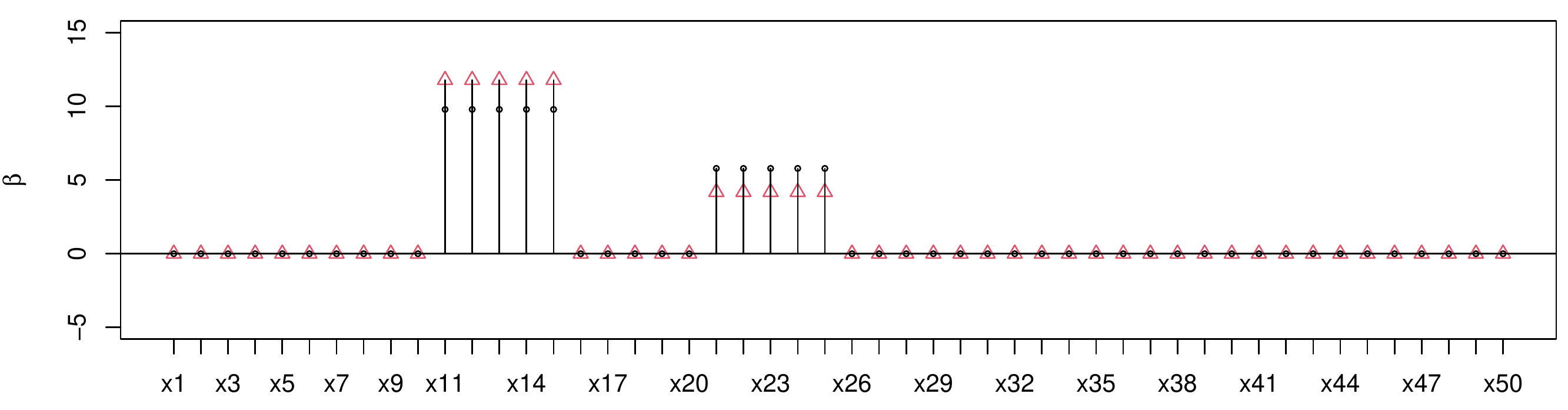}}
\label{fig:2.2}
}
\caption{Simulation results illustrating the modelling techniques. The two outcome variables are marked with $\circ$ and  $\textcolor{red}{\triangle}$.}
\label{fig:sim1.2}
\end{figure}

Next four different models are fitted to the simulated data set: Ordinary LAD regression model, LAD-lasso model and two fused LAD-lasso models with different values of $\lambda_2$. In this Section, the tuning values $\lambda_1$ and $\lambda_2$ are selected such that the differences of the examined models are pointed out clearly. Note that, in Section \ref{sec:realdata} with real-life data a standard cross-validation technique is utilized to evaluate these parameters. 
The estimates of the simulation experiment are presented in Figure
\ref{fig:sim1.2}.

Figures \ref{fig:0.0} and \ref{fig:2.0} show the effect of lasso penalty on the regression coefficients. Using tuning value $\lambda_1=0.2$ the smallest coefficients are shrunk toward zero. This type of shrinkage is a tool for variable selection in the model, which is especially valuable in a scenario where large number of the true regression coefficients are near zero. On the other hand, we can see in \ref{fig:2.0} that lasso penalty does not encourage sparsity in differences between adjacent coefficients: within the two groups of correlated  explaining variables the variation of the estimated coefficients is quite large. 

Figures \ref{fig:2.1} and \ref{fig:2.2} show the effect of a fusion penalty to the 
objective function. When the fusion penalty with tuning parameter $\lambda_2=0.1$ is used, 
the coefficient vectors $\bs\beta_5$ and $\bs\beta_{40}$ are shrunk toward zero and within the two groups of correlated  explaining variables the variation was considerably decreased (see Figure \ref{fig:2.1}). When the value of the fusion tuning parameter is increased to $\lambda_2=0.2$, the variation within the two groups of correlated explaining variables decreases to zero (see Figure \ref{fig:2.2}). In this scenario, the fused LAD-lasso models were able to both simplify the estimated model and find the two correlated covariate groups by smoothing the estimated coefficients. As a downside, the heavy smoothing somewhat underestimates the coefficient vectors $\bs\beta_5$ and $\bs\beta_{40}$.

%\section{Real-life data example: Disability retirement analysis}
\section{Application in disability retirement analysis}
\label{sec:realdata}

In this section we use the presented fused methodology to analyze retirement behaviour. The data consist of disability pension retirees in Finnish statutory earnings-related pension system for the public sector employees (both genders together). A random sample of 4,069 ($n \gg p$) municipal employees is drawn from the comprehensive administrative registers. The study-design includes three outcomes: length of working life (years) by the end of 2016, amount of final pension (Eur/month in 2017 price level) in 2017, and age at retirement (years) in 2017. The explaining variables include: monthly pensionable wages in 1995-2017 (x1-x22) and long-term sickness benefit days in 2010-2017 (x23-x30), both measured yearly. The consecutive wage observations are highly correlated (see Figure \ref{fig:wages-correlation}), whereas among benefit days the correlation is relatively mild (see Figure \ref{fig:sickness-correlation}). The design allows us to illustrate the block-structure in the covariates and robustness of the estimates in the multi-outcome situation.

\begin{figure}[ht!]
\centering
\subfloat[Asinh of earnings-related pension]{%
\resizebox*{6cm}{!}{\includegraphics[scale = 0.3]{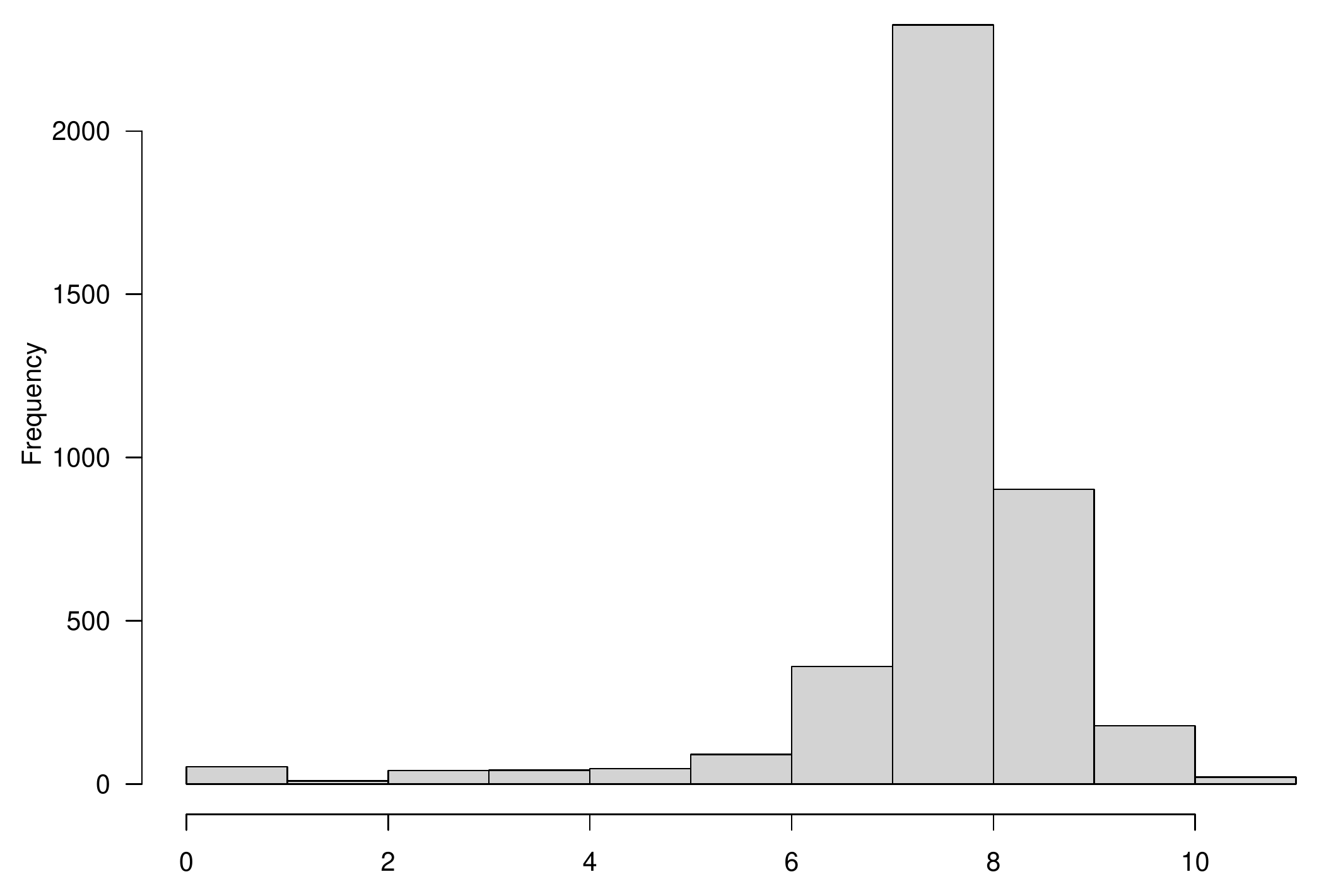}}}
\\
\subfloat[Length of working-life]{%
\resizebox*{6cm}{!}{\includegraphics[scale = 0.3]{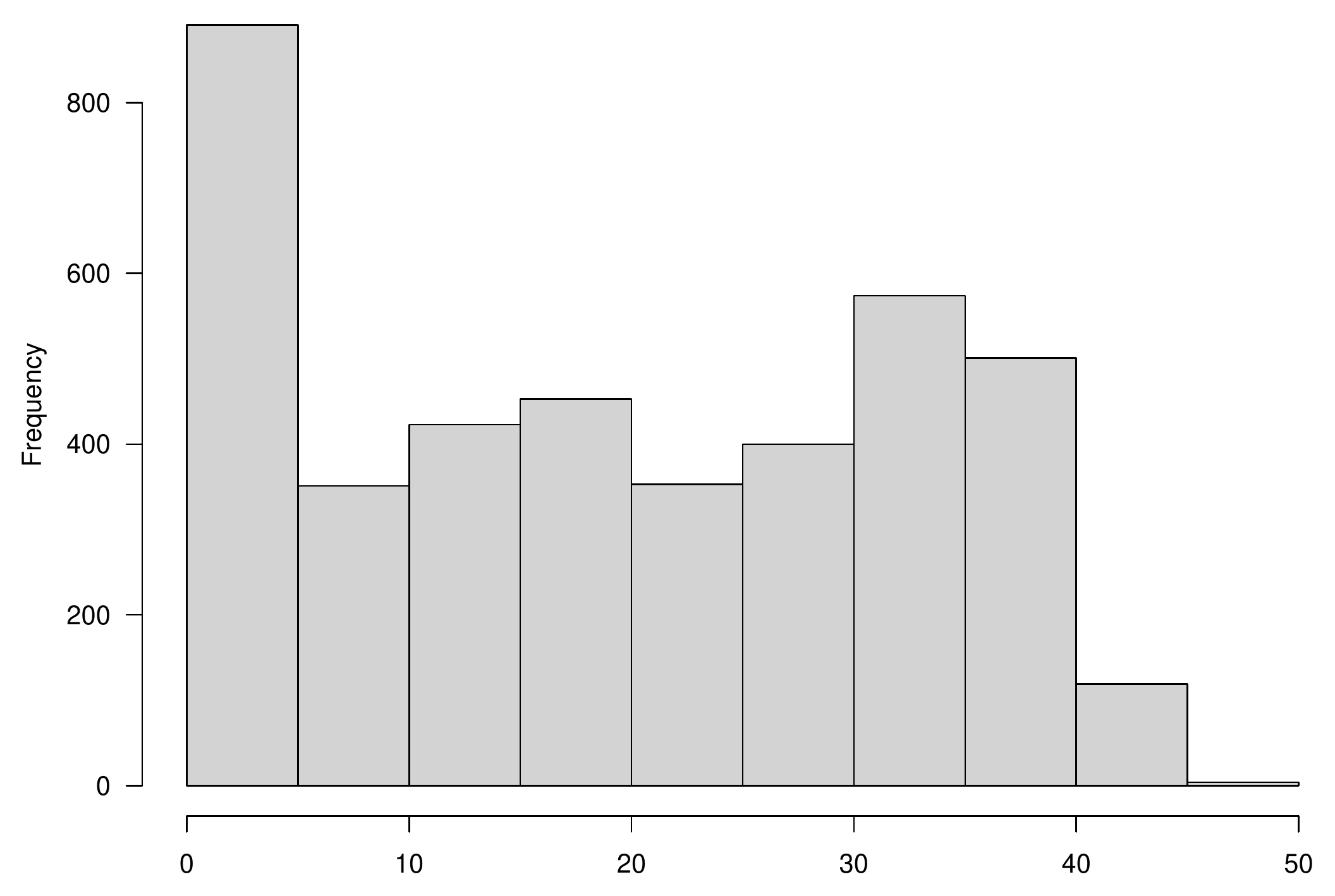}}}
\\
\subfloat[Retirement age]{%
\resizebox*{6cm}{!}{\includegraphics[scale = 0.3]{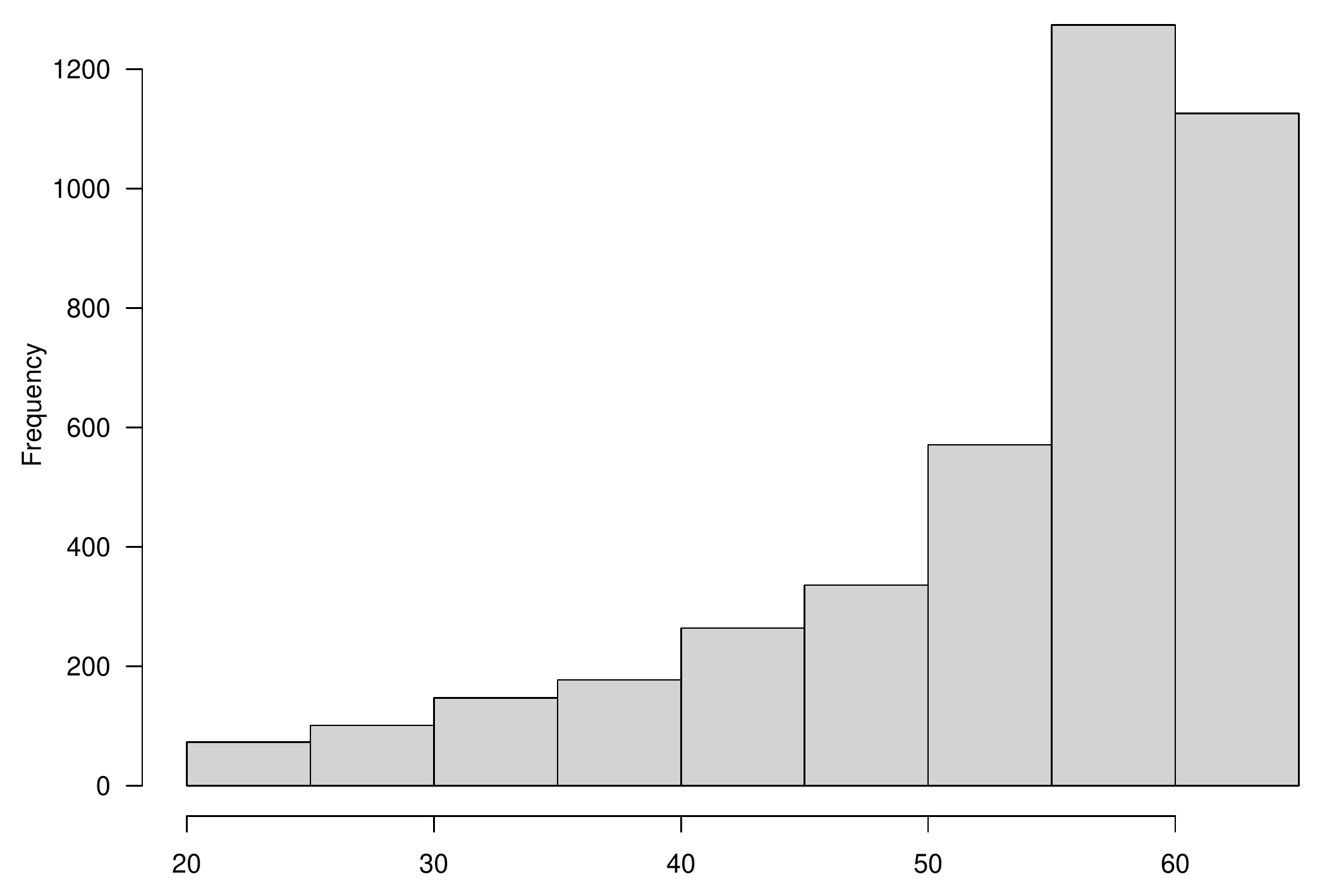}}}
\caption{Histograms of outcome variables: asinh earnings-related pension (a), length of working life (b) and retirement age (c).}
\label{fig:outcomes}
\end{figure}

\begin{figure}[ht!]
\centering
\subfloat[Monthly wages]{%
\resizebox*{14cm}{!}{\includegraphics[scale = 0.6]{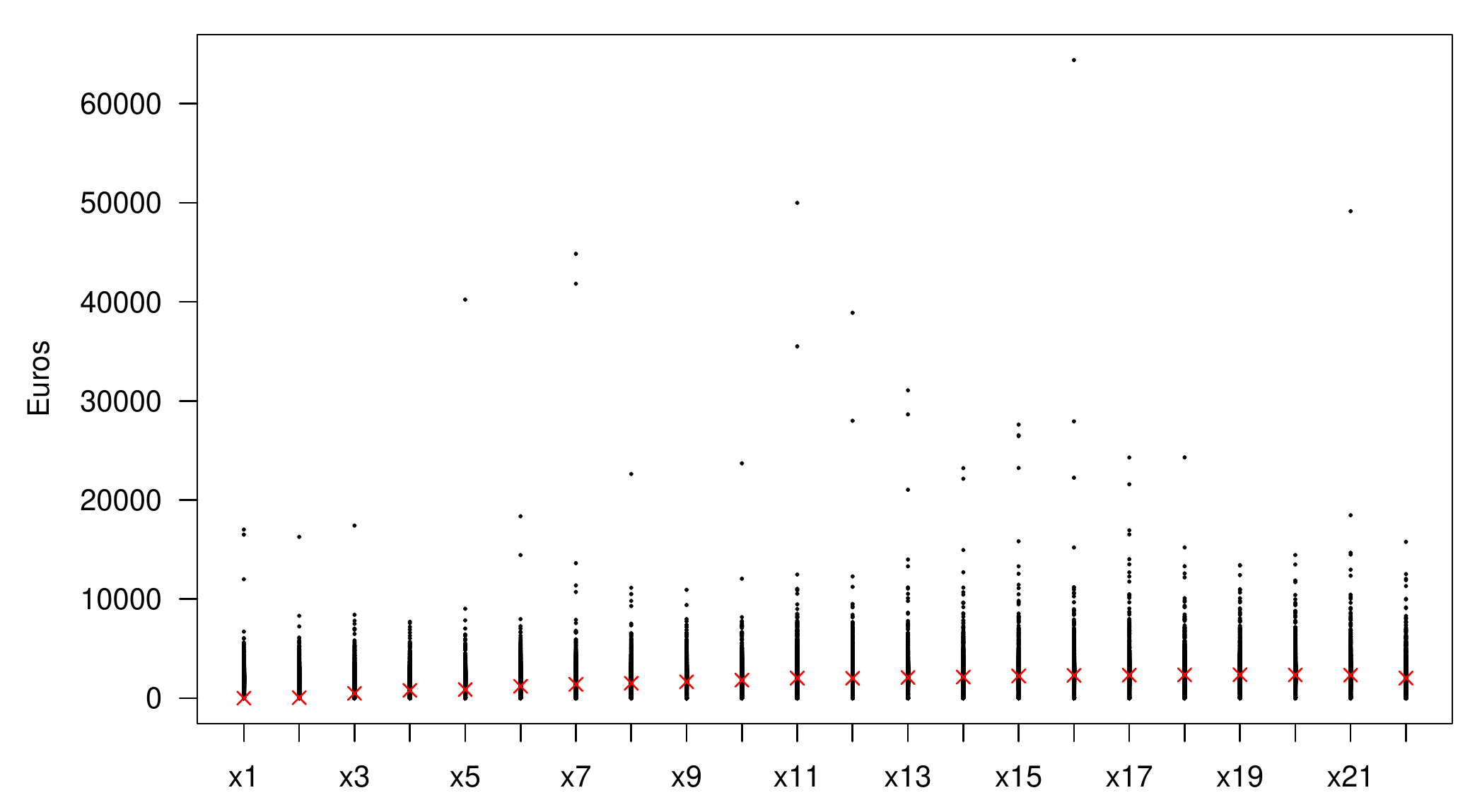}}}
\\
\subfloat[Long-term sickness benefit days]{%
\resizebox*{14cm}{!}{\includegraphics[scale = 0.6]{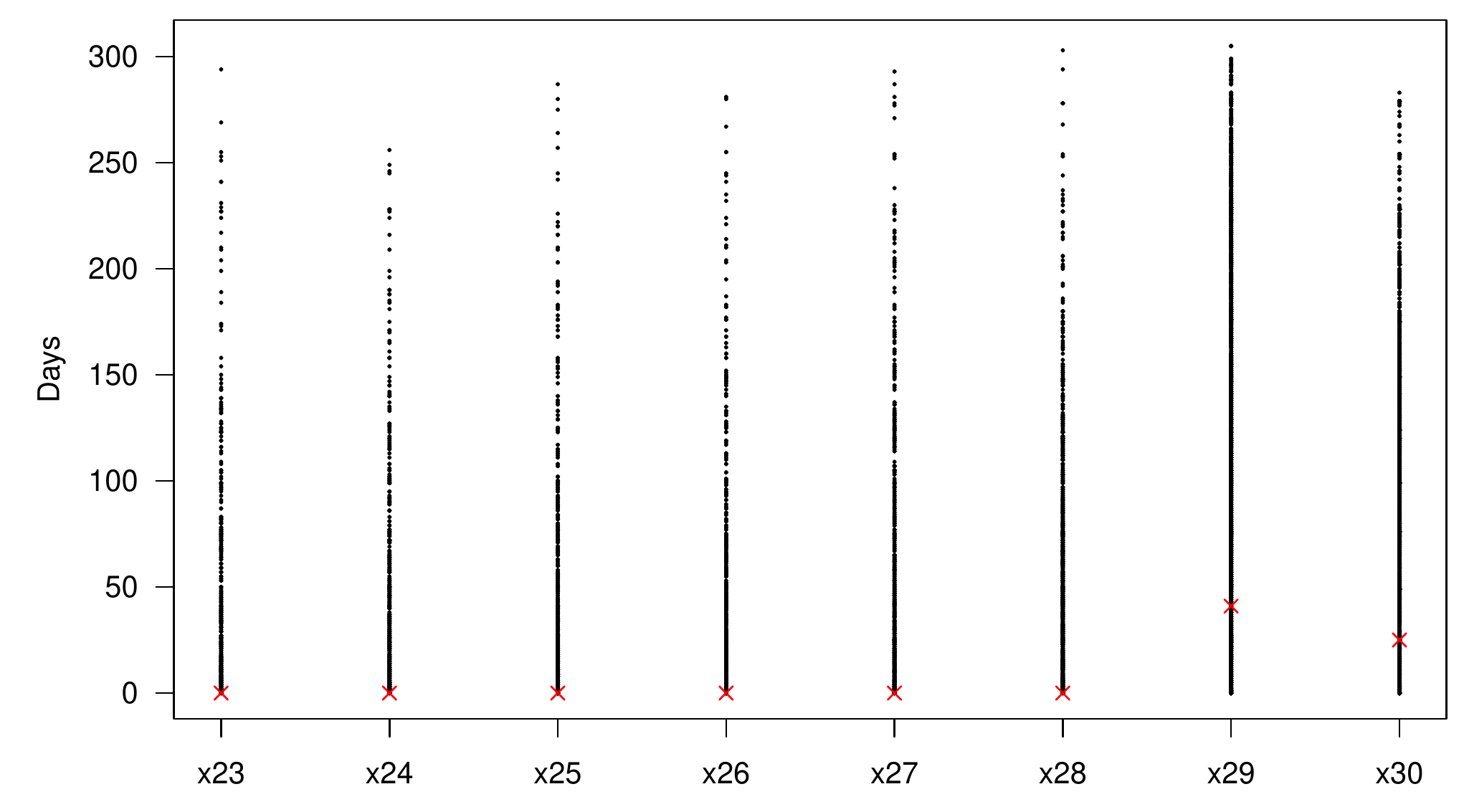}}}
\caption{Scatter plots of explaining variables: monthly wages (a) and long-term sickness benefit days (b). Median is marked with \textcolor{red}{$\times$}.}
\label{fig:predictors}
\end{figure}

The distributions of the outcomes have long tails, both left and right, and they also contain zeros (see Figure \ref{fig:outcomes}). The distributions of explaining variables have long right tails and excess number of zeros, which have a real-life relevance (see Figure \ref{fig:predictors}). In the subsequent analysis wages and pensions are transformed using \textit{asinh} transformation (e.g.,\cite{Bellamare2020}). Long-term sickness is oftentimes associated with deteriorating work ability and may lead to permanent disability, but in many cases work ability is lost without prolonged illness, and therefore we observe excess number of zeros in the measure. A note is in order on the fact that in practice zero yearly wages during life-course follow from absence from work for different reasons, such as, working in private sector or as self-employed, or drawing other social security benefits (e.g., unemployment, parental leaves). From Figure \ref{fig:predictors} we can also see that the wage distribution contains some outlying observations which challenge the existing methods.

\begin{figure}[ht!]
\centering
\subfloat[LAD estimates ($\lambda_1=\lambda_2=0$).]{%
\resizebox*{14cm}{!}{\includegraphics[width=1\textwidth]{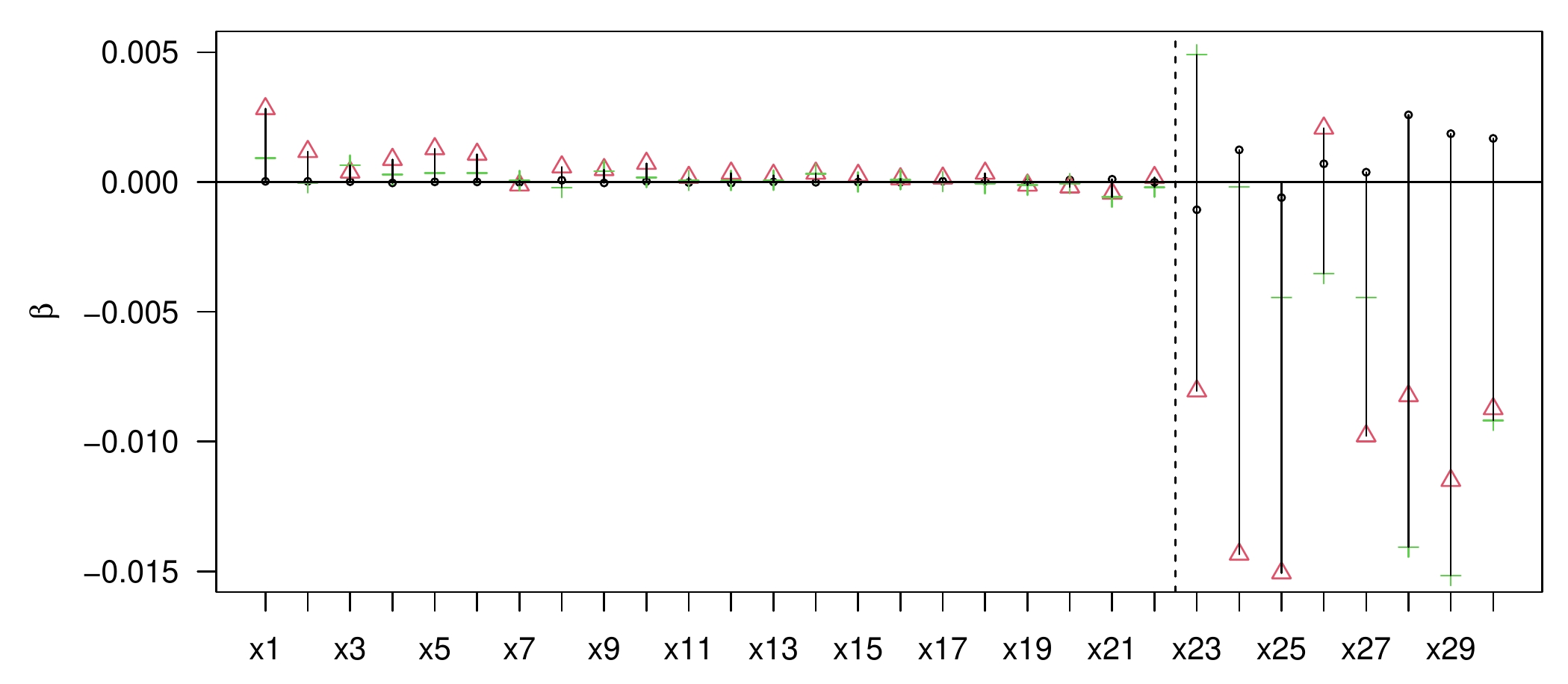}}
\label{fig:LAD}
}\\
\subfloat[Fused LAD-lasso estimates ($\lambda_1=0.03, \lambda_2=9.13$).]{%
\resizebox*{14cm}{!}{\includegraphics[width=1\textwidth]{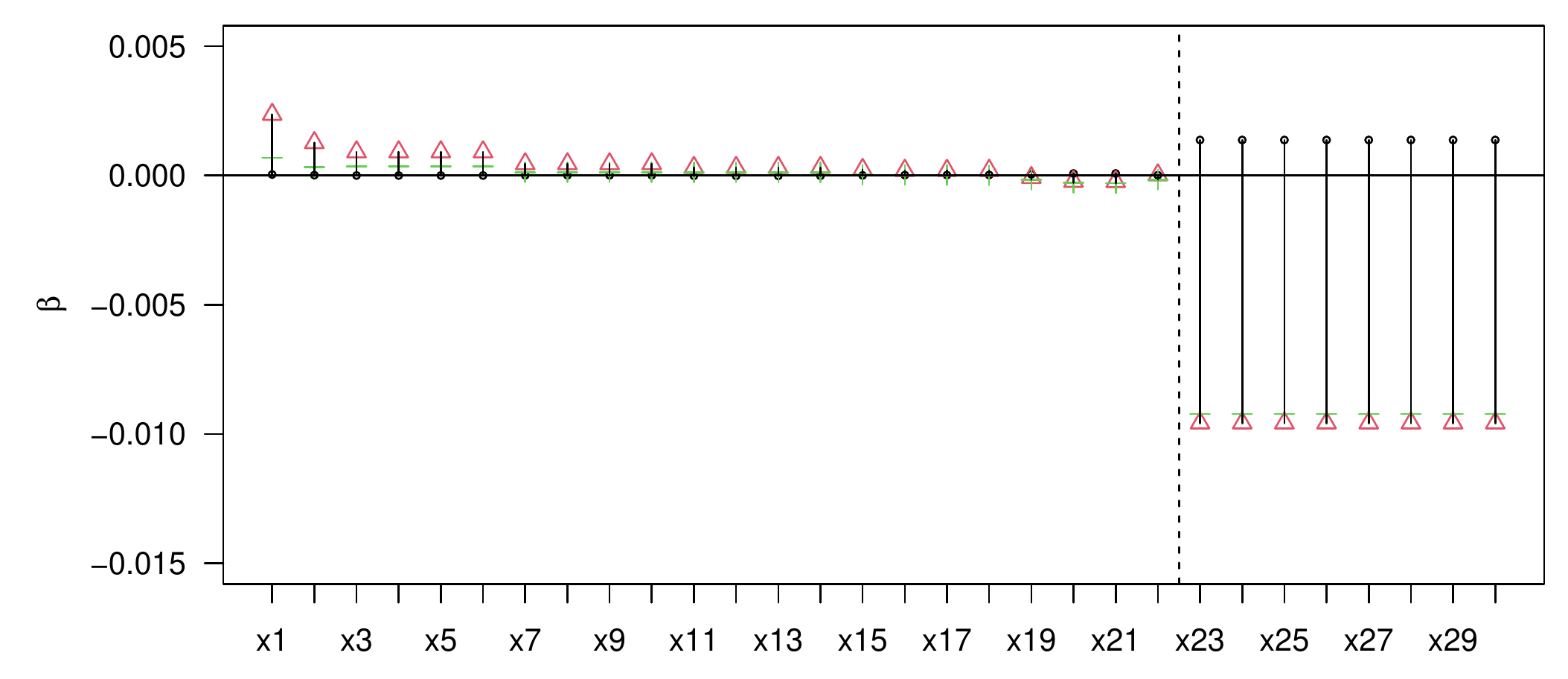}}
\label{fig:Fused_LAD-lasso}
}
\caption{LAD-estimates and fused LAD-lasso estimates. Outcome markers: length of working life ($\textcolor{red}{\triangle}$), retirement age ($\textcolor{green}{+}$) and amount of pension ($\circ$). Explaining variables: asinh monthly wages  in 1995-2017 (x1-x22) and  long-term sickness days in 2010-2017 (x23-x30).}
\label{fig:image3}
\end{figure}

From Figures \ref{fig:LAD} we can see the coefficient estimates for ordinary LAD-lasso without the fusion penalty, and fused LAD-lasso for the three outcomes. In fused LAD-lasso the covariate blocks $\gamma_j$ are defined for wages (x1-x22) and benefit days (x23-x30).

The ordinary LAD regression estimates for wages indicate higher values for distant wages especially for outcomes: length of working life and retirement age. In benefit days we see high variation across outcomes and time-points.

Next, we switch to fused LAD-lasso estimation results. First step of the analysis is to evaluate the $\lambda_1$ and $\lambda_2$ parameters. Since there are currently no alternative criterion for the model selection, we used the \textit{5-fold cross-validation criterion} (CV criterion) similar to \cite{mottonen2021}. As a search procedure we use \textit{grid search}, which is a common and easy to implement model selection method, where a subset of parameter values are fully searched. The contour plot of the parameter space and the minimum values of the $\lambda$'s are given in Figure \ref{fig:CV} in appendix. From Figure \ref{fig:image3} we can see the results of the fused LAD-lasso estimation with the CV criterion. The results indicate similar estimates for wages as in ordinary LAD-lasso. We can see that the fusion smooths the coefficient estimates of the outcomes, especially between distant time-points (x1-x17). The time-points for benefit days (x23-x30) are smoothed in total. We can see that length of working life and retirement age get negative estimates (-0.00957 and -0.00922), while the amount of pension gets positive estimates (0.00137). 

When comparing ordinary and fused techniques, we can see that fused technique yields lower estimates in absolute terms, which is as expected.

Finally, we can observe that there is an issue with the spike in the wage estimate (x1) for the length of working life in both analyzes. In practice, the spike in the estimate for wages in x1 (year 1995) on pension is a result of high correlation between consecutive wages. This multicollinearity in regression analysis can, in general, lead to instability in the estimates, and to a problem where the explained variance is allocated arbitrarily among the correlated explanatory variables (see \cite{Farrar1967}). To some degree the proposed fused LAD-lasso technique alleviates these phenomena, but the issue needs to be considered by the practitioners.

%\begin{figure}[ht]
%\centerline{\includegraphics[scale = 0.6]{Hist_logpension.eps}}
%\centerline{\includegraphics[scale = 0.6]{Hist_workinglife.eps}}
%\centerline{\includegraphics[scale = 0.6]{Hist_retirement.eps}}
%\caption{\label{fig:hist_logpension}Histograms of outcome variables asinh %earnings-related pension, duration of working-life and retirement age.}
%\end{figure}

%\begin{figure}[ht]
%\centering
%\includegraphics[width=1\textwidth]{Example1a.pdf}
%\caption{LAD estimates ($\lambda_1=\lambda_2=0$).}
%\label{fig:LAD}
%\end{figure}

%\begin{figure}[ht]
%\centering
%\includegraphics[width=1\textwidth]{Example1b.pdf}
%\caption{Fused LAD-lasso estimates ($\lambda_1=0.025, \lambda_2=9.13$).}
%\label{fig:Fused_LAD-lasso}
%\end{figure}

%%%%%%%%%%%%%%%%%%%%%%%%%%%%%%%%%%%%%%%%%%%%%%%%%%%%%%%%%%%%%%
%%%%%%%%%%%%%%%%%%%%%%%%%%%%%%%%%%%%%%%%%%%%%%%%%%%%%%%%%%%%%%
\section{Conclusion and discussion}
\label{sec:concluding}
%We introduced the fused LAD-lasso into the multi-outcome context and showed its performance with simulation and a real-life data example.
We presented the multi-outcome fused LAD-lasso, which introduces the robust LAD-lasso regression with a fusion penalty into a multi-outcome context. In a simulation study, we pointed out the pros of the proposed method to both achieving variable selection and finding the true correlated groups in the sequential data. In a disability retirement study, we applied the method to a challenging real-life data set, with outlier observations and a skewed distribution of outcome variables. The proposed method provided insights to the analysis by smoothing the highly correlated neighboring covariates.  We believe that practitioners in many fields of science familiar with regression techniques could find the fused LAD-lasso modelling approach useful when robust estimates are required. 

A note is in order on the fact that in high-dimensional data the lasso-estimated non-zero coefficients may still contain some false positives. To filter out false positives, \textit{stability selection} has been proposed as an additional confirmatory step after lasso-estimated variable selection \citep{meinshausen2010}. This kind of confirmatory step is also known as \textit{post-selection inference}; see a recent review on the topic \citep{zhangym2022}. This step was omitted here but it may be beneficial to include it in the analysis if the consistency of the variable selection is one of the main requirements.

In the disability retirement study, the evaluation of optimal tuning parameters $\lambda_1$ and $\lambda_2$ was accomplished with a CV criterion using grid search to sift through the possible values of $\lambda_1$ and $\lambda_2$.  However, when the models are complex and the number of observations is high, applying more advanced methods such as \textit{Bayesian optimization} would be beneficial in terms of computational time \citep{Shahriari2016}. In essence, Bayesian optimization constructs a probabilistic model for the objective function, which is then iteratively updated based on the evaluations with different values of hyperparameters ($\lambda_1$ and $\lambda_2$ in our case), while taking into account the uncertainty of the model \citep{Snoek2012}. Although the overall computational time in our analysis was reasonable, the use of Bayesian optimization instead of CV may be necessary with larger data sets in the future.

The LAD-lasso modelling approach, in general, would benefit from further development. First, the model selection criterion (to select appropriate $\lambda_1$ and $\lambda_2$ values) naturally have influence on 
%%affects 
the results. 
%%somewhat. 
Currently, the CV criterion is used, but in the future BIC-type criterion could broaden the options for model selection and validation. Second, in many empirical applications there is an issue with the excess number of zeros in explanatory variables. To alleviate the influence of zeros, an adaptive LAD-lasso has been proposed by \citep{mottonen2021} and later applied in \citep{Lahderanta2022}. The adaptive technique could be merged with the presented fused approach. Third, issues of post-selection inference and confidence intervals of the regression coefficients (cf. stability selection) in the  LAD-lasso context needs more attention in the future.

%%%%%%%%%%%%%%%%%%%%%%%%%%%%%%%%%%%%%%%%%%%%%%%%%%%%%%%%%%%%%%
%%%%%%%%%%%%%%%%%%%%%%%%%%%%%%%%%%%%%%%%%%%%%%%%%%%%%%%%%%%%%%
\section*{Acknowledgements}

%%%%%%%%%%%%%%%%%%%%%%%%%%%%%%%%%%%%%%%%%%%%%%%%%%%%%%%%%%%%%%
%%%%%%%%%%%%%%%%%%%%%%%%%%%%%%%%%%%%%%%%%%%%%%%%%%%%%%%%%%%%%%
\section*{Disclosure of interest}

The authors report no conflict of interest.

\section*{ORCID}
Jyrki Möttönen: \url{https://orcid.org/0000-0002-6270-2556}\\
Tero Lähderanta: \url{https://orcid.org/0000-0001-5856-2899}\\
Janne Salonen: \url{https://orcid.org/0000-0002-0595-6226}\\
Mikko J. Sillanpää: \url{https://orcid.org/0000-0003-2808-2768}\\

\vfill\eject

\bibliographystyle{tfs}
\bibliography{references}

\vfill\eject

\appendix
\section{Additional visualizations in the disability retirement analysis}

\begin{figure}[ht!]
\centerline{\includegraphics[scale = 0.6]{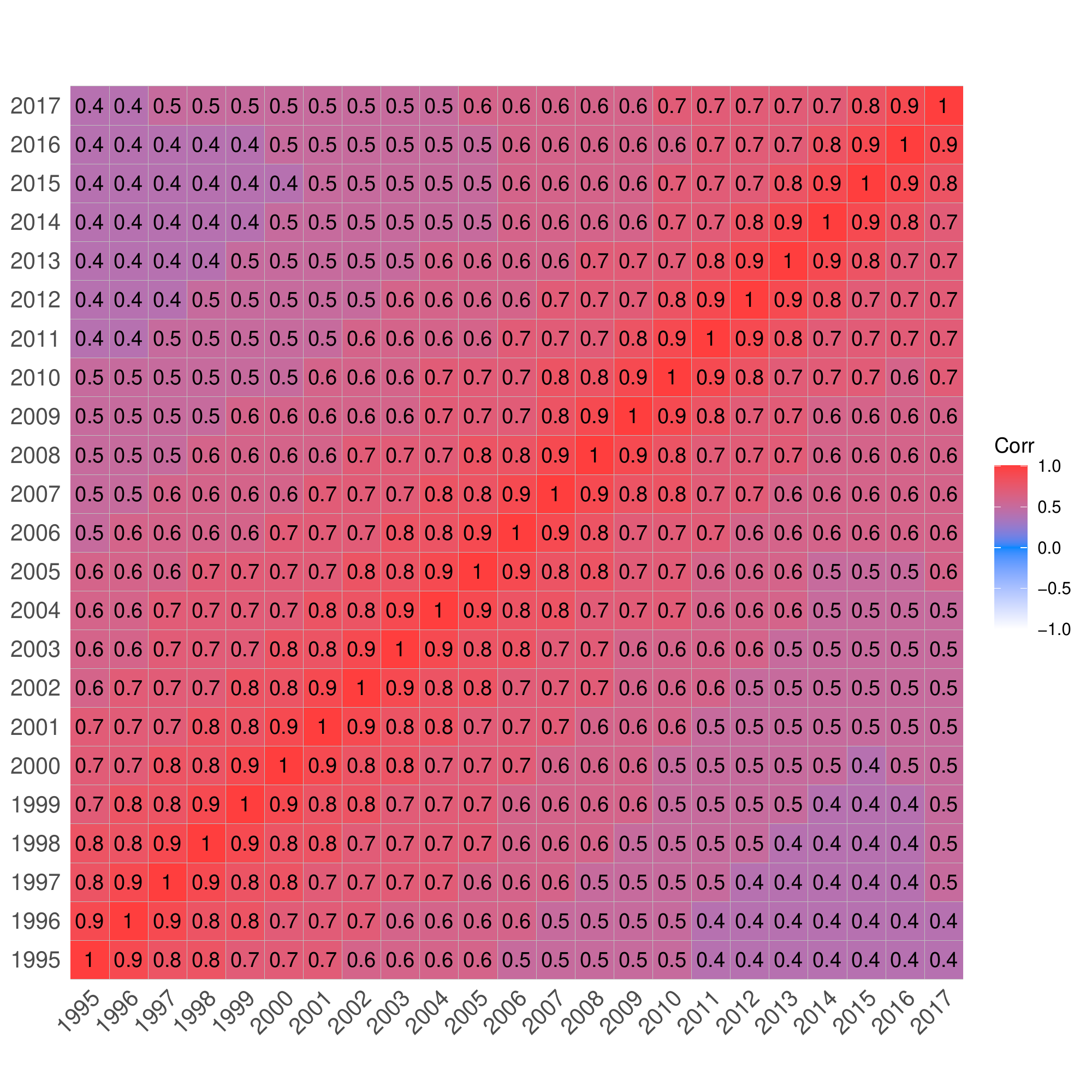}}
\caption{\label{fig:wages-correlation} The correlation plot of the wages from years 1995 to 2017 (variables x1-x22).}
\end{figure}

\begin{figure}[ht!]
\centerline{\includegraphics[scale = 0.6]{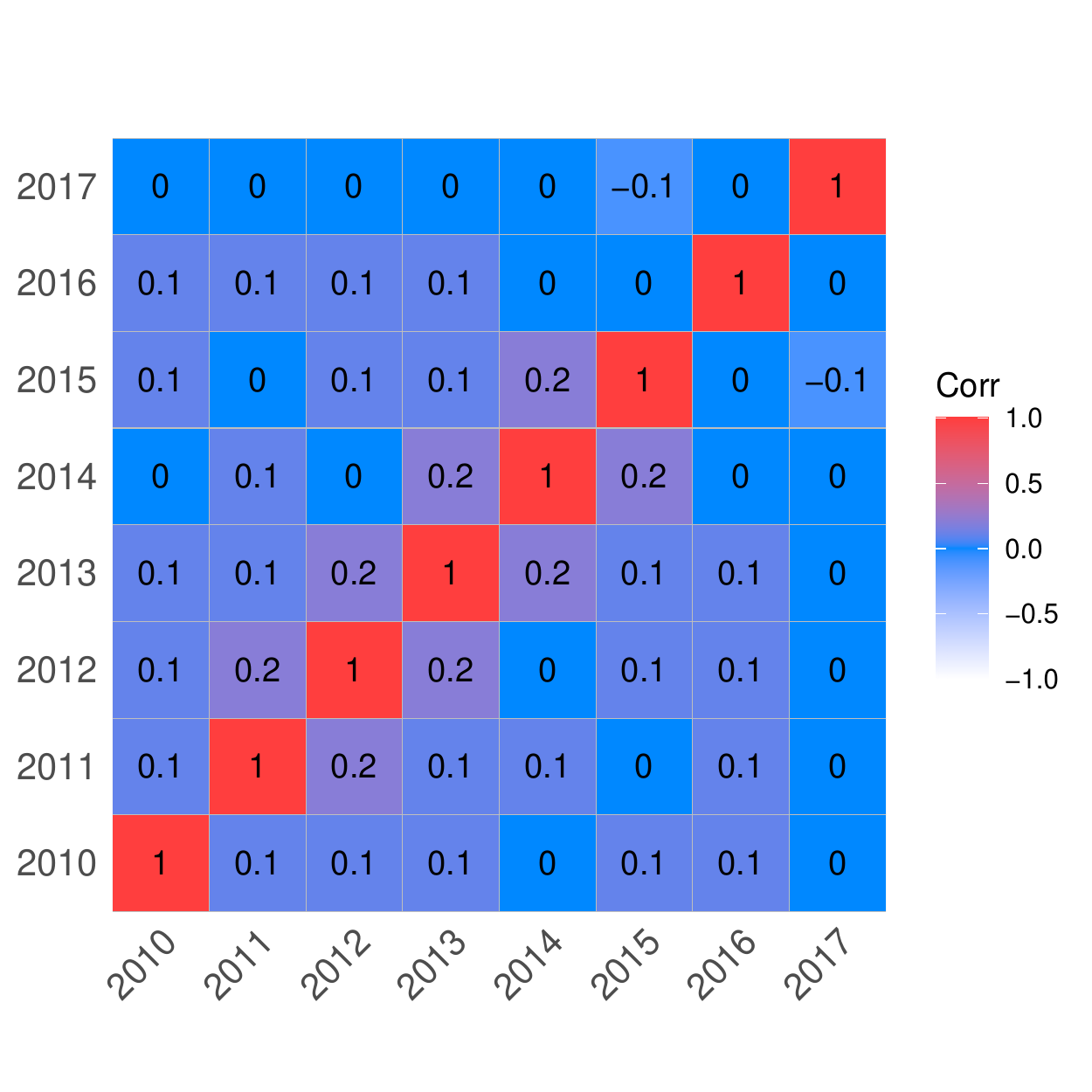}}
\caption{\label{fig:sickness-correlation} The correlation plot of the long-term sickness benefit days from years 2010 to 2017 (variables x23-x30).}
\end{figure}

\begin{figure}[ht!]
\centerline{\includegraphics[scale = 0.6]{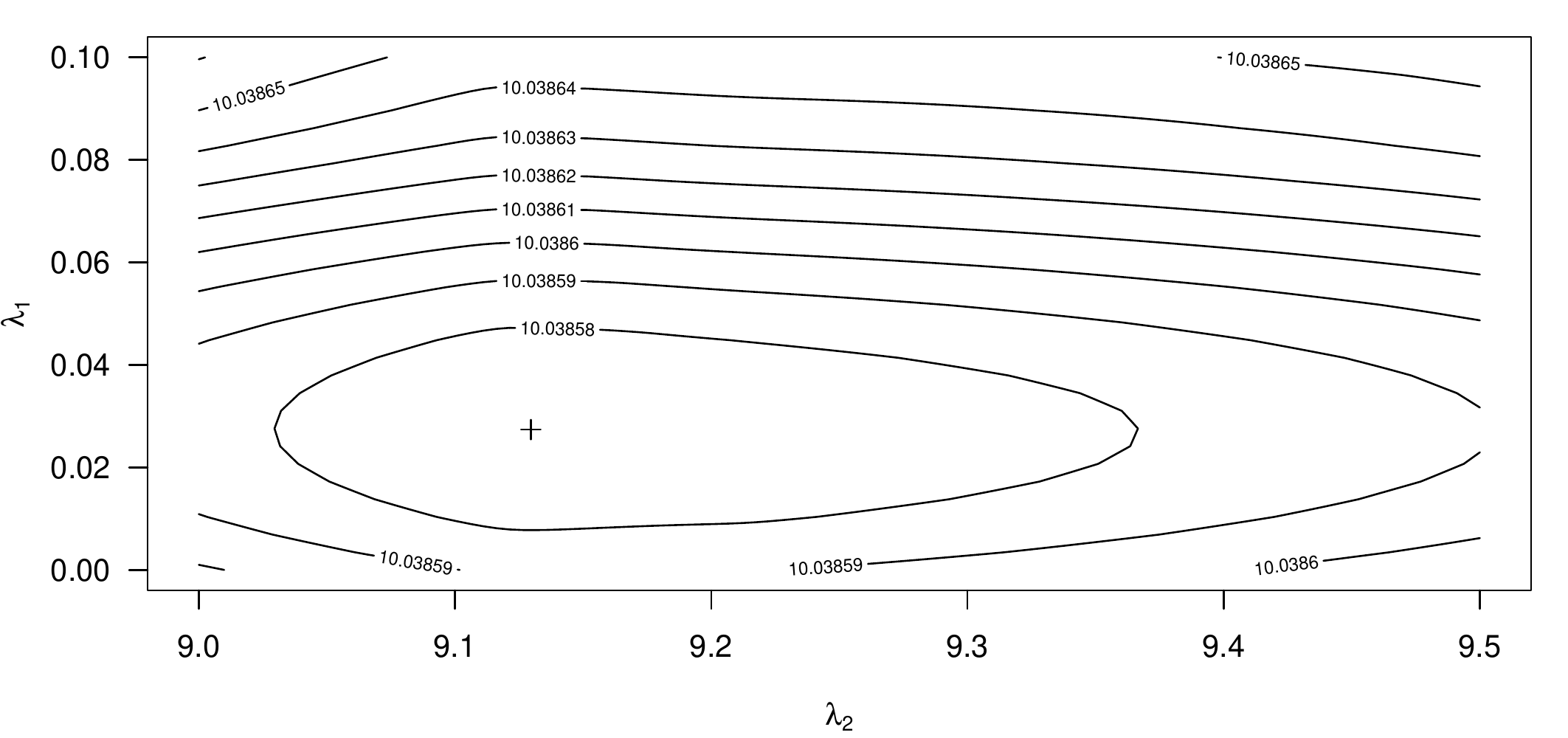}}
\caption{\label{fig:contour}Contour plot of cross-validation mean absolute errors. The global minimum point ($\lambda_1=0.03$, $\lambda_2=9.13$ ) is marked with $+$.}
\label{fig:CV}
\end{figure}
%%%%%%%%%%%%%%%%%%%%%%%%%%%%%%%%%%%%%%%%%%%%%%%%%%%%%%%%%%%%%%
%%%%%%%%%%%%%%%%%%%%%%%%%%%%%%%%%%%%%%%%%%%%%%%%%%%%%%%%%%%%%%

\end{document}